# Linking quantum mechanical features to structural phase-transformation in inorganic solids

Prashant Singh,[1,*] Anis Biswas,[1] Alexander Thayer,[1] and Yaroslav Mudryk[1]

[1]Ames National Laboratory, United States Department of Energy, Iowa State University, Ames, IA 50011, USA

**Abstract**

We present a new descriptor, i.e., local lattice distortion, to predict structural phase transformation in inorganic compounds containing lanthanides and transition metals. The descriptor utilizes local lattice and angular distortions obtained from structural optimization of experimentally known crystalline phases within state-of-the-art density-functional theory method. The predictive power of the descriptor was tested on lanthanide based $RE_2In$ (RE=rare-earth) compounds known for a variety of phase transformations. We show that the inclusion of quantum-mechanical effects through local-charge, bonding, symmetry, and electronic-structure enhances the robustness of the descriptor in predicting structural phase transformation. To gain further insights, we analyzed phononic and electronic behavior of $Y_2In$, and show that experimentally observed phase transformation can only be predicted when atomic strains are included. The descriptor was used to predict structural phase change in couple of new compounds, i.e., $(Yb_{1-x}Er_x)_2In$ and $Gd_2(In_{1-x}Al_x)$, which was validated by X-ray powder diffraction measurements. Finally, we demonstrated the generality of the proposed descriptor by predicting phase transformation behavior in different classes of compounds indicating the usefulness of our approach in mapping desired phase changes in novel functional materials.

**Introduction**

Phase transitions (PT) occurring in solid state materials are at the heart of many emerging functionalities that are based on a desired set of electrical, thermal, mechanical, optical, and magnetic properties [**1,2**] including giant magnetocaloric [**3**], magneto-resistive [**4**], magneto-strictive effects [**5**], ferroelectricity [**6**], energy-storage [**7**], shape-memory effect [**8**], polar phase-transition [**9**], piezoelectricity [**10,11**], nonlinear optical effects [**12**], large pyroelectric response [**13**], and switchable photo-responses [**14**]. Enabling many technologies, these phenomena are usually rooted in strong responsiveness of materials to various thermodynamic stimuli such as temperature (T), pressure (p), and various applied external fields (ε, E, H), and are commonly associated with structural changes involving microscopic atomic distortions [**15-17**]. In fact, the abrupt changes in materials physical properties, e.g. mechanical, magnetic, electrical, etc., are considered reliable indictors of potential structural changes, and are used in construction of phase diagrams. Vice versa, predictive manipulation of PTs is a powerful tool for engineering desired physical behaviors. Thus, uncovering the viable pathways for predictive design of PTs is very crucial for guiding the discovery of advanced energy materials [**18,19**], and to harness various functionalities [**5**]. Historically, PTs were discovered serendipitously, sometimes resulting in tragic consequences – "tin pest" being a famous example. Advances in density functional theory enabled a certain degree of prediction of a compound's stability. However, prohibitively expensive computing requirements to extract underlying structural features including group-subgroup relations and local symmetry perturbations make it difficult to develop general scheme to predict structural PTs in inorganic solids.

Despite several attempts in the past [**20-23**], we still lack a robust descriptor that connects composition and crystal structure with the physics of PT and identifies the stability of a particular crystal structure with respect to a potential phase change. The complexity of quantitative description of PT arises from a multitude of parameters defining stability of competing atomic configurations including chemical, mechanical, and/or structural features. For the accurate prediction of PT, the proposed descriptor should include key quantum mechanical information such as charge, bonding, or local symmetry in such a way that it can capture the probability of phase change within *ab initio* approaches. Thus, the primary objective of the present study is to build up a physics-informed relationship between experimentally observed crystallographic changes and theoretically predicted tendency of the atomic lattice towards distortion.

In this communication, we present a new descriptor, combining local lattice and angular distortions ($\delta_{total}^{DFT}$, described in the method section), by generalizing a perspective learned from the research on high-entropy materials [**24-30**] to predict structural PT without any *a priori* knowledge of such transition in inorganic solids. The $\delta_{total}^{DFT}$ parameter is a physics-based computationally easy to estimate metric that is based on local atomic-distortion present in the crystal structure. Before estimating $\delta_{total}^{DFT}$ parameter to assess PT behavior of inorganic solids, the experimentally known crystalline phases were carefully optimized within first principles density functional theory (DFT) method to include the quantum mechanical effect arising from change in chemistry or local atomic positions [**31,32**]. We show that structural PT in inorganic compounds can be controlled through systematic tuning of $\delta_{total}^{DFT}$ via change in chemistry. Finally, we mapped the change in $\delta_{total}^{DFT}$ on key atomic scale features such as atomic strains, bonding, hybridization, band-filling and/or electronic structure [**33**].

To exemplify, we focused on understanding the effect of varying $\delta_{total}^{DFT}$ on PT in RE$_2$In (RE=rare-earth) compounds as they exhibit variety of phase transitions [**34-40**] and, therefore, provide a fertile ground to test $\delta_{total}^{DFT}$ driven PT hypothesis. The ability of metric to predict structural PT was validated experimentally on two new pseudobinary systems, i.e., (Yb$_{1-x}$Er$_x$)$_2$In and Gd$_2$(In$_{1-x}$Al$_x$). Finally, we analyzed the $\delta_{total}^{DFT}$ of several other well-known rare-earth and transition metal compounds including Gd$_5$Si$_2$Ge$_2$ [**3**], (Mn,Fe)Ni(Si,Al) [**41**], MnFe(P,As) [**42**], R$_7$Pd$_3$ [**43,44**], and transition metal oxides to establish the generality of new descriptor in predicting PT, which are already well known for their interesting magnetofunctional properties (e.g., magnetocaloric, magnetoresistance, magnetostriction). The fundamental connection between structural features and PT found in this work will play a crucial role in designing novel functional materials for clean energy applications including environmentally-friendly solid-state cooling.

**Methods**

**Density-functional theory**: Within DFT, the valence interaction among electrons was described by a projector augmented-wave method [**31,32**] with an energy cutoff of 520 eV for the plane-wave orbitals. We used 7 × 7 × 3 (11 × 11 × 5) and 7 × 5 × 3 (11 × 9 × 7) Monkhorst-Pack *k*-mesh for Brillouin zone sampling of hexagonal (or hex) and orthorhombic (or ortho) phases of RE$_2$In compounds during structural optimization (electronic-relaxation) [**45**]. Similarly, we chose varying set of k-mesh to optimize different compounds used in **Fig. 8** depending on the type of crystal structure, which were then used to extract lattice distortion parameter. A very high accuracy of total energy and force convergence, i.e., 10$^{-8}$ eV/cell and 10$^{-6}$ eV/Å, was set as some compounds were required for phonon calculation that needs very strict convergence criteria in both energy and forces. We employ the Perdew-Burke-Ernzerhof (PBE) exchange-correlation functional in the generalized gradient approximation (GGA) [**46**]. In (semi)local functionals, such as GGA, the *f*-electrons are always delocalized due to their large self-interaction error [**47,48**]. To enforce the localization of the *f*-electrons, we perform PBE+U calculations [with a Hubbard U (5 eV; J=0.9 eV) introduced in a screened Hartree-Fock manner [**49**]. The localized *f*-electrons with negligible band-dispersion effects play an important role in determining the physical properties of rare earths materials. Despite the simplicity of implementation, the users require a clear understanding of the approximations within PBE+U functional for a precise assessment of predictive results such as electronic, magnetic, and/or PT behavior. This is due to ambiguity in determining +U values arising from its dependence on interatomic distances, atomic configurations, etc. [**50**]. To ensure the consistency of PBE+U approach throughout our work, we treated all rare-earth compounds with 4*f* electrons the same way, which is exemplified for two experimentally known systems, i.e., Gd$_2$In$_{1-x}$Al$_x$ and Eu$_2$In, where PBE+U parameters are optimized to reproduce underlying structural and magnetic behavior. We show that PBE+U provides a straightforward way to treat underestimated electronic interactions in systems with localized d-/f-electrons, which can be easily tuned to control the influence of the electronic correlations, critical to reproduce experimental observation.

**Lattice distortion calculation**: In general, the lattice distortion is defined for alloys in terms of atomic or lattice mismatch as follows (adopted from the work of Zhang *et al*. [**51**]): **δ** = (100*$\sqrt{c_i(1 - r_i/\bar{r})^2}$), where $r_i, \bar{r} (= \sum_i c_i r_i)$, and $c_i$ are atomic radii, average atomic radius, and concentration of individual elements, respectively. This "**δ**" has purely an empirical basis, which is calculated using elemental atomic radii. However, the elemental atomic radii are not correct representation of an alloying environment as it does not include realistic alloying effects such as hybridization, band-filling, or Fermi-surface effects.

To accurately capture the atomic distortion, we performed full structural relaxation (atomic and lattice) of each material class under consideration within first-principles DFT method. The magnitude of atomic displacement for a N atom system can be defined in terms of the mean square atomic displacement (MSAD) i.e.,

$$\text{MSAD} = \langle \Delta u^2 \rangle = \frac{1}{N}\sum_{i=1}^{N}\left\|r_i^{rlx}(x,y,z) - r_i^{ideal}(x_0,y_0,z_0)\right\|^2; \qquad \text{Eq. (1).}$$

where, $r_i^{rlx}(x,y,z)$ and $r_i^{ideal}(x_0,y_0,z_0)$ are the position of the $i^{th}$ atom in the "relaxed" and "ideal" lattices, respectively. The $(x,y,z)$ and $(x_0,y_0,z_0)$ are reduced coordinates of relaxed and unrelaxed (ideal, reference lattice points) atomic positions of the $i^{th}$ atom, while N is the total number of atoms. Here, the "ideal" structure is the unrelaxed structure with experimental reference lattice points, which can be low or higher symmetry structure depending on system of interest. The quantity in Eq. (1), i.e., $\left\|r_i^{rlx}(x,y,z) - r_i^{ideal}(x_0,y_0,z_0)\right\|$, is the displacement of the $i^{th}$ atom with respect to its reference or ideal lattice, which could be used for crystalline, partially-order or disorder solids [**27**]. Following this, we defined local lattice distortion (LLD) in terms of MSAD, i.e., $\langle \Delta u^2 \rangle$ as

$$\text{LLD} = \sqrt{MSAD} \text{ or } \sqrt{\frac{1}{N}\sum_{i=1}^{N}\left\|r_i^{rlx}(x,y,z) - r_i^{ideal}(x_0,y_0,z_0)\right\|^2}; \text{ Eq. (2).}$$

The MSAD can also be characterized as vector (L$_{2,1}$) norm of atomic displacements. The LLD based metric in Eq. (3), $\delta_r$, is used as a qualifier to predict structural PT in crystalline inorganic solids. The atomic positions of relaxed lattice, i.e., $r_i^{rlx}(x, y, z)$, required to estimate $\delta_r$ using Eq. (2), are extracted from fully relaxaed (cell volume, cell shape, and atomic positions) unit-cell of each inorganic solids considered in this work.

$$\delta_r = \frac{\Delta u_{x,y,z}}{\sqrt{[\Delta u_{x,y,z}]^2}} = \frac{\sum_i \sqrt{u_{ix}^2 + u_{iy}^2 + u_{iz}^2}}{\sqrt{(\sum_i u_{ix})^2 + (\sum_i u_{iy})^2 + (\sum_i u_{iz})^2}}, Eq.\,(3)$$

Till this point, we considered only positional distortion in evaluating our PT qualifier metric. However, to provide a more rigorous analytical description to our structural descriptor, we quantified angular distortions in terms of the deviations of bond angles from their ideal values in the crystal lattice. The root-mean square (RMS) angular distortion, $\delta_\theta$, can be calculated as

$$\delta_\theta = \sqrt{\frac{1}{N_{angles}} \sum_{ij,k} [\theta_{ij,k}^{rlx} - \theta_{ij,k}^{ideal}]}, \qquad \text{Eq. (4)}$$

where $\theta_{ij,k}^{rlx}$ and $\theta_{ij,k}^{ideal}$ represent the angle formed between two bonds, i-j/i-k, in the relaxed (DFT) and ideal (or reference) lattice, respectively. $N_{angles}$ is the total number of bond angles considered in the lattice, which provides a scalar measure of the overall angular deviation in the polar (or noncentrosymmetric) structures. To relate angular distortions to atomic displacements, we considered the bond vectors { $r_{ij}^{ideal}$, $r_{jk}^{ideal}$} and { $r_{ik}^{rlx}$, $r_{jk}^{rlx}$} for bonds $i - j$ and $j - k$ in the ideal and relaxed structures, respectively. Following this, the bond distortion can be defined as

$$\Delta\theta_{ij,k} = \theta_{ij,k}^{rlx} - \theta_{ij,k}^{ideal} = arcos\left[\frac{r_{ik}^{rlx} \cdot r_{jk}^{rlx}}{|r_{ik}^{rlx}||r_{jk}^{rlx}|}\right] - arcos\left[\frac{r_{ik}^{ideal} \cdot r_{jk}^{ideal}}{|r_{ik}^{ideal}||r_{jk}^{ideal}|}\right],$$

By incorporating MSAD of $\delta_r$ and $\delta_\theta$, a more comprehensive descriptor, i.e., $\delta_{total}^{DFT}$, was constructed that combines both atomic displacements and angular deviations, and written as

$$\delta_{total}^{DFT} = \sqrt{\delta_r^2 + \delta_\theta^2}, \qquad \text{Eq. (5)}$$

where, $\delta_r$ and $\delta_\theta$ are RMS of local lattice and angular distortions, respectively.

The combined descriptor captures both translational and rotational degrees of freedom in the lattice, offering a more nuanced characterization of lattice distortions. In this work, this approach is used for studying complex materials undergoing structural phase transitions, where both atomic (local lattice) distortions and angular distortions play significant roles in stabilizing various phases.

While there are many studies aimed at predicting PTs [**20-23**], to best of our knowledge, there is hardly any established descriptor similar to $\delta_{total}^{DFT}$ to classify probability of structural PT in intermetallic or partially ordered compounds. For example, in the recent work where authors studied potential ways to predict transformations in Heusler alloys [**52**] they identified 11 new compounds with anticipated structural transitions but stopped short of providing a similar descriptor. Thus, in this work, we mainly resort to direct comparison with experimental observations to benchmark our predictions of structural PT. We chose several material classes including $RE_2In$, $RE_7Pd_3$, (Mn,Fe)Ni(Si,Al), MnFe(P,As), $Gd_5Si_2Ge_2$, and transition metal oxides, to establish the accuracy of our approach.

*Uncertainty in* $\delta_{total}^{DFT}$ *determination*: Firstly, a modern experimental study can, if designed properly, unequivocally determine the presence or absence of structural change, but there is still uncertainty present about precise lattice parameters and atomic positions. The error in "ideal" structure coming from experiments may propagate in $\delta_{total}^{DFT}$ estimation while using Eq. (1-5). However, so far, we see no evidence of this uncertainty affecting the reliability of our predictions. Secondly, we note that the uncertainty in determination of atomic positions in experiments will affect any approach including, for example, symmetry mode analysis.

**Phonon calculations**: Density-functional perturbation theory (DFPT) was used to construct force constant matrix needed to calculate lattice dynamical properties [**54,55**]. The phonon dispersion plots are done along the high-symmetry direction of the Brillouin zone of crystal structure [**56-58**]. Because phonons are very sensitive to the forces on atom and cell lattices, a very high convergence criteria were set for energy and forces convergence to minimize the force on each atom as close as possible to 'zero'.

**Experimental details**: For experimental validation of our theoretical results, we prepared several $RE_2In$ samples and carried out their X-ray powder diffraction characterization at room temperature. The temperature-dependent data are taken from references [**35,38,41**]. The synthesis conditions are specified below. All the samples are polycrystalline in nature. The rare earth elements were supplied by the Materials Preparation Center of the Ames National Laboratory (>99.9 wt.% purity with respect to all elements in the Periodic Table), while In and Al (>99.99 wt. % purity) were purchased from Alfa Aesar.

*Gd$_2$(In$_{1-x}$Al$_x$) compounds*: Gd$_2$(In$_{1-x}$Al$_x$) was synthesized via conventional arc melting of appropriate amounts of Gd, Al, and In in a Zr-gettered argon atmosphere. The alloy was remelted several times, flipping the button over after each melting, to ensure homogeneous mixing. The as-cast Gd$_2$(In$_{1-x}$Al$_x$) samples were used for the present study.

*Yb$_2$In and (Yb$_{0.5}$Er$_{0.5}$)$_2$In compounds*: The Yb$_2$In and (Yb$_{0.5}$Er$_{0.5}$)$_2$In were prepared using induction furnace. At first, the constituent elements were sealed in Ta- crucibles under a partial atmosphere of ultrapure Ar. Then, crucibles containing samples were placed inside high frequency induction furnace and melting was done for one hour at 1473 K for Yb$_2$In and 1873 K for (Yb$_{0.5}$Er$_{0.5}$)$_2$In. To ensure homogeneous mixing, the alloys were remelted at the same conditions after flipping the crucibles. After induction melting, the crucibles were sealed in Ar-filled quartz tubes and placed inside a resistive furnace for annealing at 873 K for two weeks.

**Results and discussion**

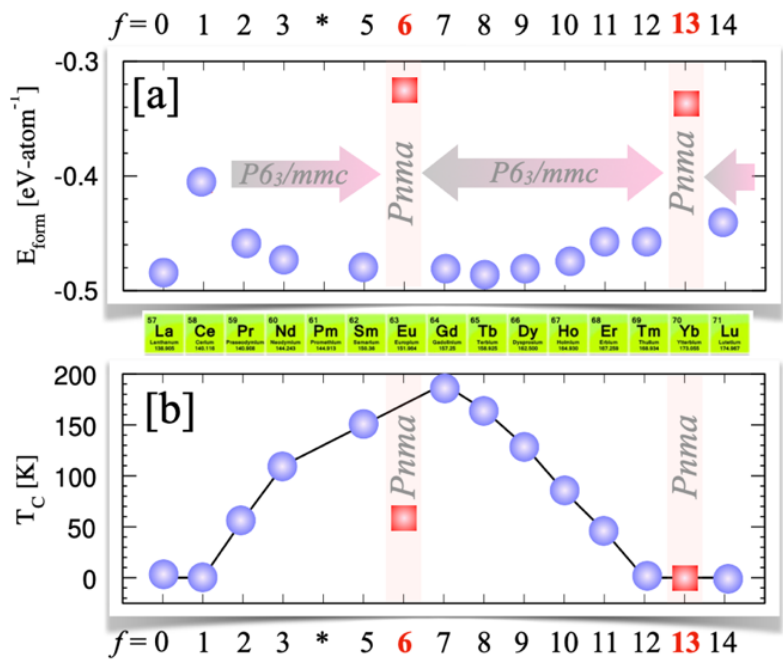

**Figure 1**. (a) DFT calculated formation energy (E$_{form}$ (0K); eV-atom$^{-1}$) of RE$_2$In compounds (RE = La-Lu). (b) The experimental Curie temperature (T$_C$; K) of RE$_2$In in their respective ground state, i.e., *P6$_3$/mmc* (sg=194) and *Pnma* (sg=62). The *P6$_3$/mmc* and *Pnma* phases of RE$_2$In compounds are marked by circle (solid blue) and square (solid red), respectively.

In rare earth compounds, the relationships between the various physical and thermodynamic properties such as phase stability and Curie temperature take place according to *4f* electron count, which changes as we move across the lanthanide series in the Periodic Table. We would like to recall the case of RE$_2$In (RE= rare-earth) compounds for which the formation of orthorhombic (*Pnma* space group) ground state is stabilized for only two alloys with RE= Eu and Yb as shown in **Fig. 1a** while the rest of the members of the series possess the hexagonal (space group *P6$_3$/mmc*) ground state.

Firstly, we analyzed the trends in DFT calculated formation energies (E$_{form}$, **Fig. 1a**) and experimental Curie-temperature (T$_C$, **Fig. 1b**) for RE$_2$In compounds across the lanthanide series. The filling of 4*f* sublevels (also see **SI Fig. 1** that shows similar trends in mechanical properties and Hund's spin arrangement) produces a jump in E$_{form}$ (0 K) at Eu and Yb (**Fig. 1a**), which clearly reflects electron transfer from 5$d^1$ states to 4$f^6$/4$f^{13}$ leading to 4$f^7$/4$f^{14}$ configuration. Notably, these electron fillings lead to structural transformation from hexagonal (*P6$_3$/mmc*) phase to orthorhombic (*Pnma*) phase for Eu$_2$In and Yb$_2$In compounds. Similar trend was also observed in magnetic ordering temperature, which leads to anomalous T$_C$ in Eu$_2$In (**Fig. 1b**). Interestingly, a short-range magnetic anomaly exists in Eu$_2$In around 150 K, which follows the general Curie temperature trend for Eu$_2$In [**35**]. This shows that change in electron-count or charge density has a significant role to play in controlling crystallography and physical properties. One should also keep in mind that divalent Eu and Yb atoms are substantially larger than neighboring trivalent RE atoms, leading to potentially larger local distortion. This atomic size difference stems mainly from the change in *f*-electron configuration, further magnifying its importance. In addition to lanthanide elements (i.e., La-Lu), we also note that RE$_2$In compounds also form with 3*d* and 4*d* rare earth elements such as Sc and Y. The comparative analysis of thermodynamic stability (i.e., formation energy) of Y-In, Eu-In, and Yb-In (**Fig. 2a-c**) indicates that the orthorhombic phase of Y$_2$In, Eu$_2$In, and Yb$_2$In compounds fall on the convex hull line confirming the thermodynamic stability for all three cases, which is in good agreement with the existing literature [**34-37**]. Interestingly, the hexagonal (*P6$_3$/mmc*) phase of Y$_2$In observed at high temperature was found unstable at lower temperatures, which is in good agreement with experiments [**34**].

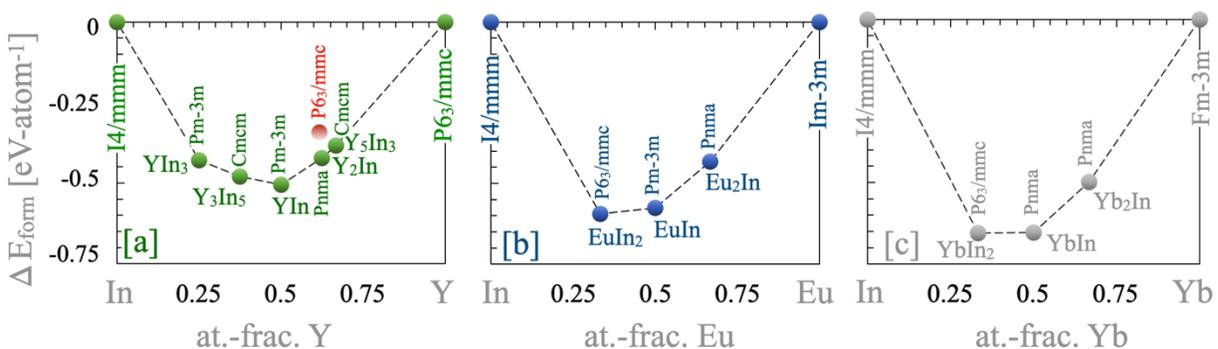

**Figure 2**. The thermodynamic phase stability of (a) Y$_2$In, (b) Eu$_2$In, and (c) Yb$_2$In compounds on their respective convex hull diagrams confirming orthorhombic phase as a ground state structure.

To provide a fundamental basis, we present a schematic showing connection between lattice distortion and body-centered cubic (bcc) to hexagonal PT in **Fig. 3**. In an earlier study, Burgers et al. [**47**] have pointed to martensitic distortion (bcc → hcp) driven PT without shearing, and independent of its glide modes. The atomic distortion (martensitic) can be obtained by considering an orthorhombic vectors close to bcc and hcp lattices (see **Fig. 3a,b**), which was also discussed by Bowles *et al.* and others in Ref. [**59-61**]. For example, distortion matrix of bcc→ hcp transformation can be calculated by considering the non-orthogonal frame constituted by the normalized bcc axes x = $(1/\sqrt{3})[111]_{bcc}$, y = $(1/\sqrt{3})[11\bar{1}]_{bcc}$, and z = $(1/\sqrt{2})[\bar{1}10]_{bcc}$. This way the distortion in bcc phase can transform $[110]_{bcc}$ plane defined by $[11\bar{1}]_{bcc}$ and $[111]_{bcc}$ into the $[0001]_{hcp}$ plane without significant cell deformation via local atomic displacements along {111} direction. For easy visualization, the distortion planes in hcp phase are shown in **Fig. 3b** highlighting fundamental correlation of atomic distortions across the crystal phases, e.g., bcc and hcp. These distortions sometime may include several structural aspects including tilting and rotations of local crystal motifs (e.g., octahedra, tetrahedra, etc.) along with bond-length/angle variations.

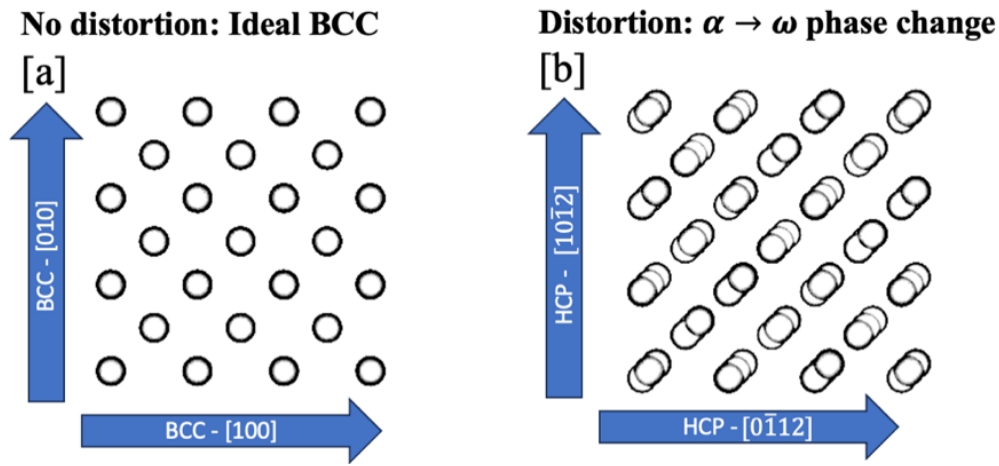

**Figure 3**. A simplified schematic showing degree of martensitic distortion in the symmetrically equivalent planes for (a) BCC to (b) HCP phase change.

The structural phase transition from hexagonal Ni$_2$In/ZrBeSi (*P6$_3$/mmc*) to orthorhombic Co$_2$Si/TiNiSi (*Pnma*) structure or vis-a-vis is a common occurrence in transition metal-based compounds, e.g. of M'M''X composition (M is a *d*- and X is a *p*-element) that gives rise to intriguing magneto-functional properties [**35,43**]. Notably, the orthorhombic phase is considered a distorted version of the hexagonal phase where the lattice parameters in two phases are closely related by symmetry as *a$_{orth}$* ≡ *c$_{hex}$*, *b$_{orth}$* ≡ *a$_{hex}$*, and *c$_{orth}$* ≡ √3*a$_{hex}$*, which is similar to bcc to hcp martensitic PT as shown in **Fig. 3**. Now a relevant question would be that how to establish correlation between $\delta_{total}^{DFT}$ and its connection to PT.

To test our $\delta_{total}^{DFT}$ hypothesis, we chose Y$_2$In, which is a RE$_2$In type compound without *4f* electrons. As shown in **Fig. 2a**, Y$_2$In stabilizes with orthorhombic phase in its ground state and exhibits a temperature-driven structural PT from low-T orthorhombic phase to high-T hexagonal phase [**34**], therefore, it provides an opportunity to study connection between $\delta_{total}^{DFT}$ and structural PT. Our calculations show that local-lattice distortion for Y$_2$In in the hexagonal phase is very small ($\delta_{total}^{DFT}$~ 0) while orthorhombic phase shows relatively large distortion ($\delta_{total}^{DFT}$~0.97; $\delta_r$~0.93). Therefore, to gain deeper insights, we analyzed the bonding, charge, and phonon behavior of fully relaxed hexagonal and orthorhombic phases of Y$_2$In to interpret their relations with the atomic distortion, which is then compared with similar features of the orthorhombic phase with "hypothetical reference lattice" (**Fig. 4**). Here, by "hypothetical reference lattice", we mean that the atoms in orthorhombic unit-cell are placed at point-symmetry elements, i.e., special Wyckoff positions. We have highlighted the difference between "ideal", "relaxed", and "hypothetical reference lattice" in **SI Table 2**. The DFT-relaxed hexagonal and orthorhombic Y$_2$In structures show two (3.361, 3.120 Å) and four (3.166 – 3.593 Å) distinct bonds, respectively, while the *Pnma* phase with "hypothetical reference lattice" has almost identical Y-Y, Y-In, and In-In bond-lengths (see **SI Fig. 4**, and **SI Table 3**).

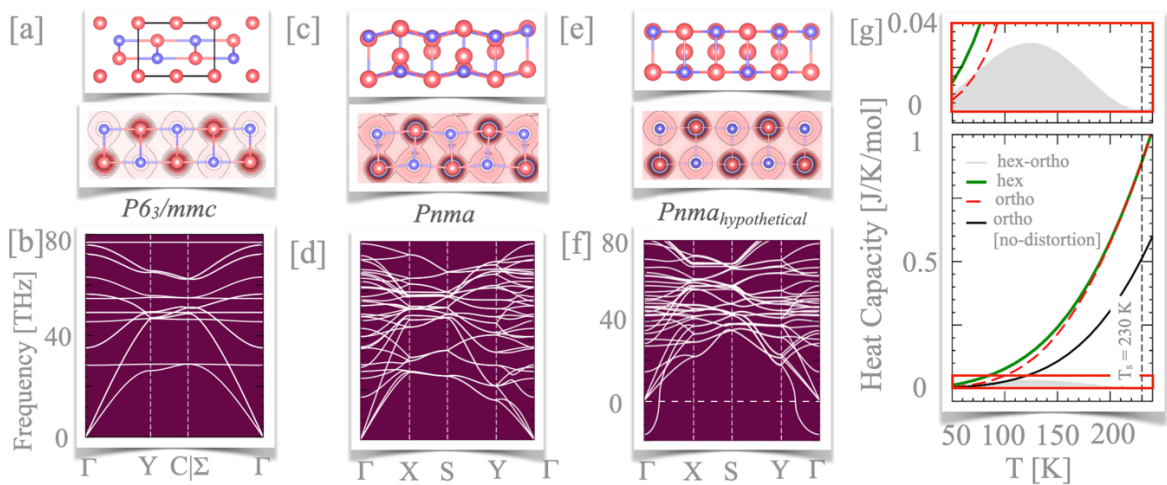

**Figure 4**. The three Y$_2$In lattice arrangements: (a, b) *P6$_3$/mmc,* (c, d) *Pnma*, and (e, f) *Pnma* ("hypothetical reference lattice"), showing change in degree of distortion and its effect on atomic charge density and phonon dispersion. (g) The specific heat shows crossover between *P6$_3$/mmc* and *Pnma* crystal phases at 230 K, in good agreement with experiments (250±5K) [**35**]. The shaded area in (g) and zoomed region (top-panel) shows difference in specific heat between two phases.

As shown in **Fig. 4a**, both bond and structural anisotropy (Y-In bond anisotropy >7%) in hexagonal phase provides enough charge for neighboring sites, as shown by directional distribution in 2D projected charge density plots, ensuring energy stability (**SI Fig. 3**), which explains why hexagonal phase does not show any local distortions (**Fig. 4a**). The positive phonons (**Fig. 4b**) also confirm the structural stability of hexagonal phase of $Y_2In$ compound. Unlike hexagonal phase, the fully relaxed orthorhombic (*Pnma*) phase shows strong atomic and bond distortion that is reflected in distorted charge density around atomic sites (**Fig. 4c**). The structural stability of relaxed orthogonal phase is confirmed by phonons showing no imaginary modes (**Fig. 4d**).

To probe the distortion hypothesis, we created an orthorhombic "*hypothetical reference lattice*" (*Pnma*) where Y and In in $Y_2In$ are placed at special "high-symmetry" positions, i.e., Wyckoff points (**Fig. 4e**). The projected charge density shows nearly spherical distribution in **Fig. 4e**, which indicates that the lack of atomic distortion prohibits any kind of charge transfer required for phase stability. Despite these local re-arrangements, the average Y-In bond length in relaxed (3.360 Å) vs experimental (3.369 Å) *Pnma* phase remains almost same (less than 0.3% change). Thus, the origin of structural instability in transition metal-based compounds has been attributed to either a cooperative Jahn-Teller distortion (with local moments) or band Jahn-Teller distortion (without local moments) [28]. Notably, our results also show that local distortion improves the structural stability of $Y_2In$ in orthorhombic phase (**Fig. 4d**) while undistorted lattice remains dynamically unstable at 0 K (**Fig. 4f**). We attribute this dynamic stability of distorted (**Fig. 4c**) orthorhombic lattice to higher degree of charge sharing between neighboring atoms compared to orthorhombic "hypothetical reference lattice" (**Fig. 4e**). Clearly, the local atomic distortion modifies the underlying charge sharing mechanism in different $Y_2In$ crystal phases, which greatly controls the structural stability.

Interestingly, the phonon dispersion in undistorted orthorhombic setting shows an imaginary mode near the Γ-point confirming structural instability (**Fig. 4f**). We also compared the specific heat of three cases in **Fig. 4g** where hexagonal and realistic orthorhombic phases show a clear crossover at 230 K, in a good agreement with experimentally observed PT at 250±5K [36]. The zoomed region (top-panel) in **Fig. 4g** is shown to highlight specific-heat crossover at the transition temperature between the two phases. On the other hand, specific heat of undistorted orthorhombic phase (black curve **Fig. 4g**) is clearly distinct from other two phases. This model analysis explains the need for necessary distortion in the orthorhombic phase to supply enough charges to neighboring sites to ensure phase and structural stability. To understand the reason for weaker stability of hex phase low temperatures, we analyzed the electronic structure of $Y_2In$ in hexagonal and orthorhombic phases.

*Electronic-structure connection to structural stability:* In **Fig. 5a-c$_1$**, we show electronic-structure evidence for a structural PT in $Y_2In$ from the high-temperature hexagonal (*P6$_3$/mmc*) to a low-temperature orthorhombic (*Pnma*) phase. Similar to negative phonons that suggest structural instability (e.g., **Fig. 4f**), the presence of a peak in the density of states at the Fermi level is often correlated to possible phase instability, see Ref. [**62,63**]. While looking at equivalent high-symmetry point (i.e., $\Gamma$) in both *P6$_3$/mmc* (**Fig. 5a**) and *Pnma* phase with "hypothetical reference lattice" (**Fig. 5b**), we can clearly see a flattened degenerate band where a flat band produces a peak in DOS (also see Fermi-surface with flat bands in **Fig. 5b$_1$**) near/at Fermi level in an undistorted Y$_2$In. The Fermi surface was also found to show interesting topology, where distorted cylinders centered at the Z-point in *Pnma* phase with hypothetical reference lattice show very small electron-hole pockets (**Fig. 5b$_1$**). On the other hand, electronic-structure (**Fig. 5c**) and the Fermi-surface of distorted *Pnma* phase (**Fig. 5c$_1$**) show much larger electron-hole pockets at Z-point, i.e., reduced instability.

Notably, the band-structure analysis shows that a change in phase stability could indeed be attributed to Jahn-Teller type distortion that lifts of band degeneracy due to increased distortion. The Y-In bond anisotropy contributes to structural instability (**Fig. 4f**) by creating inequivalence between *Y* and *In* atoms. This symmetry breaking results into splitting of degenerate bands along $\Gamma$-Z-U (**Fig. 5c**) due to electronic charge redistribution driven local atomic distortion, which lowers the overall energy and stabilizes the orthorhombic phase (see **Fig. 4d**). This observation can be also understood in terms of Jahn-Teller model where structural phase transition often occurs in compounds with a sharp *d*-electron peak (high-density band) near/at the Fermi level [**34**]. Similar behavior was also noted in YCu [**64**] and LaCd [**65**]. Finally, the orthorhombic phase with no atomic distortion shows imaginary (unstable) phonon modes (**Fig. 4f**) with band degeneracy along $\Gamma$-Z and Z-U part of electronic-structure in **Fig. 5b**. This shows that local atomic distortion is needed to stabilize orthorhombic phase of Y$_2$In, highlighting direct correlation with our $\delta_{total}^{DFT}$ driven PT hypothesis. This suggests that quantitative assessment of $\delta_{total}^{DFT}$ can possibly be used to distinguish orthorhombic and hexagonal PT where higher $\delta_{total}^{DFT}$ would show preference for orthorhombic phase while lower $\delta_{total}^{DFT}$ would favor hexagonal phase.

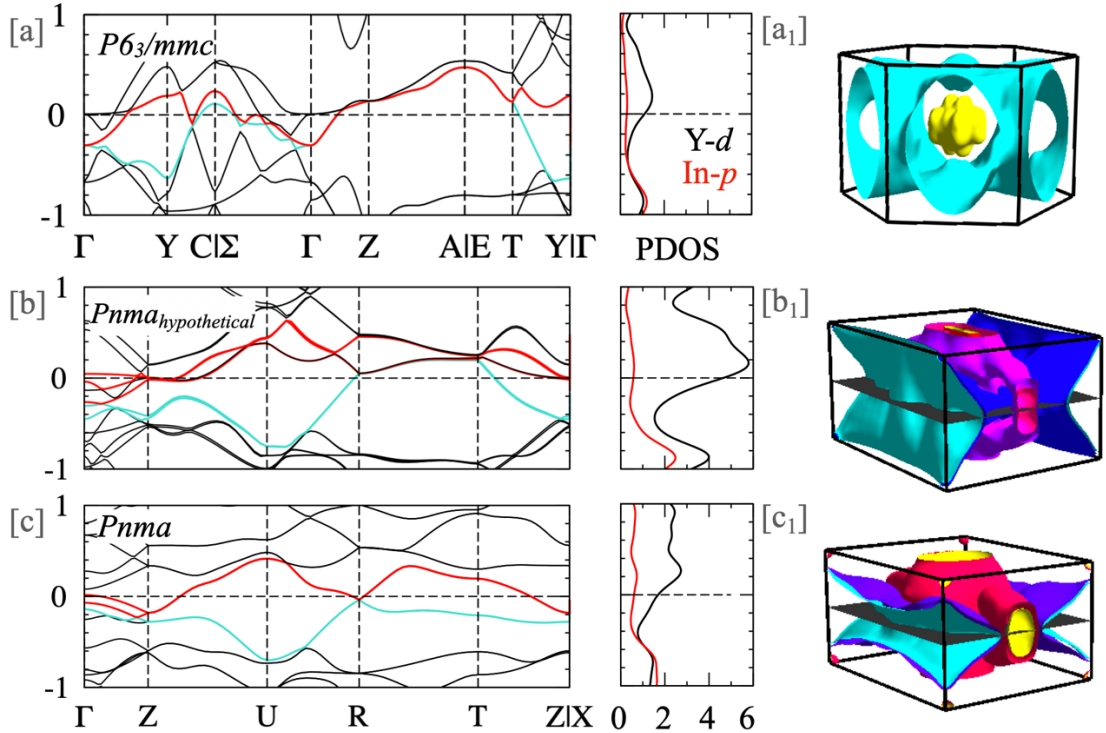

**Figure 5**. Evolution in electronic structure and Fermi surface with local structural changes in (a, a$_1$) *P6$_3$/mmc*, (b, b$_1$) *Pnma ("hypothetical reference lattice")*, and (c, c$_1$) *Pnma* phases of Y$_2$In compound. Notably, a change in phase stability could indeed be attributed to Jahn-Teller type distortion that lifts the band degeneracy due to increased distortion.

*Distortion controlled structural and magnetic phase change exemplified through (Yb$_{1-x}$Er$_x$)$_2$In and Eu$_2$In*: In the previous section, we showed a direct connection between $\delta_{total}^{DFT}$ and PT in Y$_2$In compound and discussed the underlying reason why Y$_2$In has orthorhombic ground state. Moreover, our calculations also support strong $\delta_{total}^{DFT}$ change in going from high temperature hexagonal ($\delta_{total}^{DFT} \sim 0$) phase to low temperature orthorhombic ($\delta_{total}^{DFT} \sim 1.0$) phase of Y$_2$In compound.

In this section, we focus on two new examples, i.e., (Yb$_{1-x}$Er$_x$)$_2$In (**Fig. 6**) and Eu$_2$In (**Fig. 7**), to test and validate the predictive ability of the proposed metric. The choice of (Yb$_{1-x}$Er$_x$)$_2$In system was based on the fact that Yb$_2$In is orthorhombic throughout the whole temperature range (due to high distortion **SI Fig. 5**) [37], while Er$_2$In is hexagonal, which makes it interesting to see if we could modify their existing ground states by chemical substitution of divalent Yb by trivalent Er. Previous experimental study [37] confirmed that Yb for Eu substitution in the (Yb$_{1-x}$Eu$_x$)$_2$In system, where both rare earths are divalent, does not change crystal structure, i.e. it remains orthorhombic.

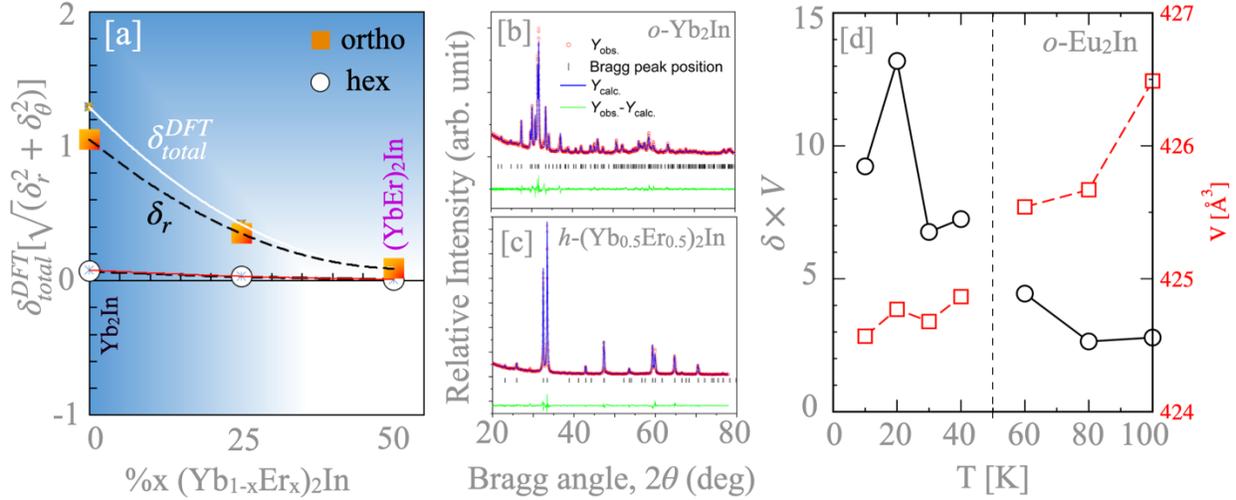

**Figure 6**. (a) The reduction in calculated $\delta_{total}^{DFT}$ (solid white line) and $\delta_r$ (dashed black line), predicting orthogonal to hexagonal phase change in $(Yb_{1-x}Er_x)_2In$, with increasing Er concentration. (b, c) X-ray powder diffraction patterns confirming formation of the orthorhombic phase for $Yb_2In$ at high $\delta_{total}^{DFT}$, and hexagonal phase for $(Yb_{0.5}Er_{0.5})_2In$ at low $\delta_{total}^{DFT}$, validating our predictions. (d) In orthorhombic $Eu_2In$, the lattice distortion vs temperature (circle, in black) plot shows a sharp change near magnetic transition temperature coinciding with a jump in unit cell volume (square, in red). The magnitude of $\delta_{total}^{DFT}$ change from the magnetic transition is smaller than depicted, but, for the visual reasons, we scaled the $\delta_{total}^{DFT}$ with the experimentally determined cell volume [**35**].

**Figure 6a** shows that the magnitude of $\delta_{total}^{DFT}$ in the orthorhombic phase of $(Yb_{1-x}Er_x)_2In$ reduces significantly with increasing Er concentration, which is indicative of decreased orthorhombic character. Notably, the $\delta_{total}^{DFT}$ parameter in $(Yb_{1-x}Er_x)_2In$ at x=0.5 (Er) diminishes to zero, and if we look back at our discussion of hex-to-ortho PT in $Y_2In$, the low $\delta_{total}^{DFT}$ in $(Yb_{0.5}Er_{0.5})_2In$ suggests that hexagonal phase should be more stable phase compared to orthorhombic phase. By synthesizing these compounds and analyzing their crystal structure by X-ray powder diffraction measurements shown in **Fig. 6b** (x=0.0) and **Fig. 6c** (x=0.5), we were able to validate our predictions that low $\delta_{total}^{DFT}$ indeed stabilizes the hexagonal phase while high $\delta_{total}^{DFT}$ preferably favors orthorhombic phase ($Yb_2In$).

Similar to $Yb_2In$, $Eu_2In$ also forms an orthorhombic phase and shows no structural/symmetry change [**26**]. However, $Eu_2In$ shows a magnetic PT in association with symmetry invariant discontinuous change in volume at 57 K (see **Fig. 6d**) [**35**]. To understand how $\delta_{total}^{DFT}$ analysis connects with the first-order magnetic PT in $Eu_2In$, we show $\delta_{total}^{DFT}$ associated with lattice modulations at experimental volume in **Fig. 6d**. Notably, the $\delta_{total}^{DFT}$ shows a temperature dependent change in going from 60 K ($\delta_{total}^{DFT}$ =4.44 in units of cell volume; V=35.46 Å³/atom) to 40 K ($\delta_{total}^{DFT}$ =7.24 in units of cell volume; V= 35.40 Å³/atom) at the magnetic crossover point. Notably, the DFT calculated $\delta_{total}^{DFT}$ shows sharp increase near/at the magnetic ordering temperature of 57 K as observed in unit cell volume ($\Delta V/V \cong +0.1\%$) and lattice constant ($\Delta a/a \cong 0.08\%$, $\Delta b/b \cong 0.04\%$ and $\Delta c/c \cong 0.03\%$) [**35**]. This further suggests that the knowledge of $\delta_{total}^{DFT}$ can be utilized to qualitatively map the discontinuous first-order magnetoelastic transition even for the case when the transition is associated with infinitesimal change in lattice parameters or volume [**66-69**]. Finally, we want to caution the user that this metric was developed with the intent to predict structural phase transition but interestingly, we were able to capture the change in $\delta_{total}^{DFT}$ arising from magnetic transition, which is exemplified for Eu$_2$In. However, at this stage, it is difficult to claim the generality of this approach in predicting magnetic transition as we didn't analyze enough material classes, which is not in the scope of this work.

*Distortion driven prediction of PT in Gd-In-Al compounds*: Next, we take Gd$_2$(In$_{1-x}$Al$_x$) as a mirror example, where the parent phase, i.e., Gd$_2$In, shows hexagonal crystal structure. **Figure 7a** shows that aluminum doping increases $\delta_{total}^{DFT}$ parameter, therefore, promoting the formation of orthorhombic phase. We found that the distortion-induced structural transformation from hexagonal to orthorhombic phase is highly probable as the concentration of Al goes from 0 to 50 at% due to the systematic increase of $\delta_{total}^{DFT}$. As shown in **Fig. 6a**, this transformation is also well captured by our distortion parameter.

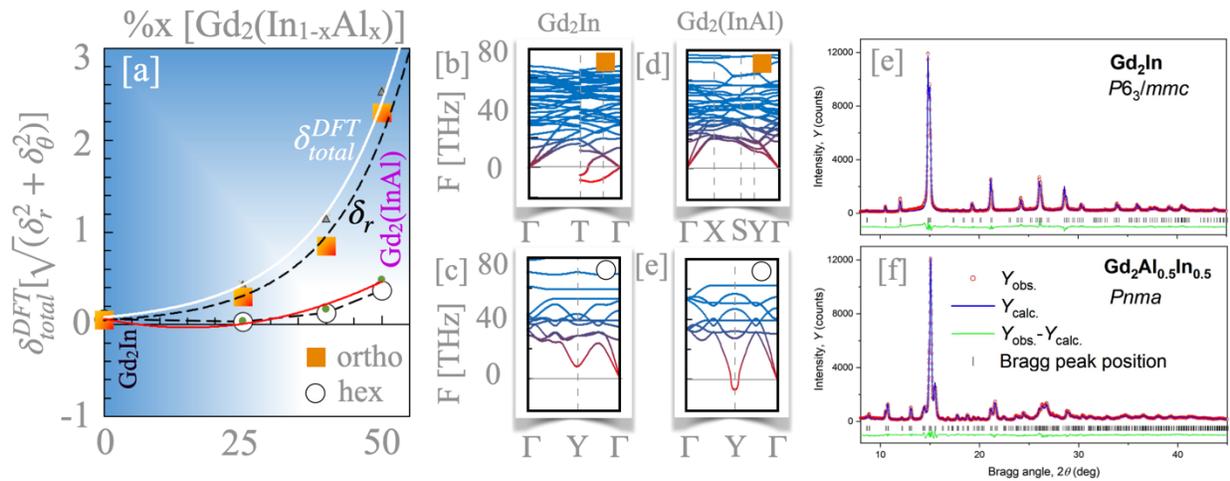

**Figure 7**. (a) The compositional dependence of $\delta_{total}^{DFT}$ (solid white line) and $\delta_r$ (dashed black line), hexagonal and orthorhombic phases of Gd$_2$(In$_{1-x}$Al$_x$) compound. (b-e) The phonon dispersion shows structural stability for Gd$_2$In and Gd$_2$(In$_{0.5}$Al$_{0.5}$) phases in hexagonal (*P6$_3$/mmc*) and orthorhombic (*Pnma*) phases, respectively. Both orthorhombic phase of Gd$_2$In and hexagonal phase of Gd$_2$(In$_{0.5}$Al$_{0.5}$) show negative phonons that suggest structural instability. (e,f) The X-ray powder diffraction patterns validates $\delta_{total}^{DFT}$ predicted structural phase stability in Gd$_2$In (ortho → hex) and Gd$_2$(In$_{0.5}$Al$_{0.5}$) (hex → ortho) compounds.

To further assess the structural stability of predicted phases, in **Fig. 7b-e**, we show phonon dispersion of Gd$_2$(In$_{1-x}$Al$_x$) at x=0 and 50 at.% in hexagonal (*P6$_3$/mmc*) and orthorhombic (*Pnma*) phases. The hexagonal phase was found stable at x=0 (Gd$_2$In, see **Fig. 7c**), showing good agreement with the experimental data [**29,69**], and our ionhypothesis (**Fig. 2&3**) that materials with low $\delta_{total}^{DFT}$ prefer hexagonal phase as a ground state. Furthermore, we want to emphasize that the inability of *In* to draw enough charge in *Al*-rich case is the main reason for the instability of hexagonal phase, while high $\delta_{total}^{DFT}$ = 2.5 at x=50 at.% (Gd$_2$In$_{0.5}$Al$_{0.5}$; see **Fig. 7d**) favors the orthorhombic PT as validated by our X-ray powder diffraction measurements in **Fig. 7f.**

*Structural connection between hexagonal and orthorhombic phases*: The RE$_2$In compounds adopt one of the two popular structure types, either hexagonal Ni$_2$In or orthorhombic Co$_2$Si, depending on type of rare-earth element. The vast number of intermetallic compounds known today adopt these closely related structure types – the orthorhombic phase is a distorted version of the hexagonal one – and sometimes exhibit polymorphic transformations between these two atomic structures in response to external stimuli. Therefore, in this section, we want to expound on the lattice relationship between hexagonal (*P6$_3$/mmc, hex*) and orthorhombic (*Pnma, ortho*) phases.

As mentioned earlier, the lattice of *hex* phase ($a_{hex}$, $c_{hex}$) is related to the orthorhombic phase ($a_{ortho}$, $b_{ortho}$, $c_{ortho}$) as $c_{hex} \rightarrow a_{ortho}$, $a_{hex} \rightarrow b_{ortho}$ and $3a_{hex} \rightarrow c_{ortho}$, see more details regarding *hex-to-ortho* structural transformation by Wayman *et al.* [**70**]. The above lattice relationship between two phases shows a shrinkage of the y-axis and an expansion along the x-axis with a negligible change in the z-axis direction. Using this information, we could establish rectangular coordination, i.e., X=[0001]$_{hex}$ //[100]$_{ortho}$, Y = [2110]$_{hex}$ // [010]$_{ortho}$, and Z= [0110]$_{hex}$ // [001]$_{ortho}$ (see **Ref. 70** and **SI Eq. 1** for more details regarding this structural transformation). The y and z axes are two invariant lines creating a habit plane that can easily rotate during the transformation; however, it also presents an existing energy barrier required for scuffling of certain planes in order to achieve *hex-to-ortho* PT (see **Ref. 70**). The higher distortion level of the orthorhombic structure provides enough charge and bonding strength to stabilize it over the *hex*-phase, which indicates the ways to understand the important role of $\delta_{total}^{DFT}$ in structural PT, similar to the case of bcc→ hcp PT (see related discussion in **Fig. 3**).

*Distortion predictions in other compounds*: To better understand the general concept of lattice distortion and its influence on the structural stability and transformation behavior, we have systematically calculated the effect of chemistry change on the degree of distortion for a range of crystalline solids. In **Fig. 8a**, we plot the $\delta_r$ for several broad classes of rare-earth and transition metal based functional materials with hexagonal or orthorhombic ground state crystal structure. We categorized $\delta_r$ range as follows: (a) low distortion region ($\delta_r$<0.15) with hexagonal phase as a ground state, (b) mid distortion range (0.15<$\delta_r$<1.00) that shows temperature-induced *ortho-hex* PT (low symmetry to high symmetry phase), and (c) high distortion region ($\delta_r$>1.0) with orthorhombic phase only. The atomic distortion vs displacement plot shows direct correlation with structural phase change behavior for several material classes. For clarity, we have tabulated $\delta_r$ values of RE$_2$In compounds in **SI Table 1**.

Notably, the distortion parameters for the RE$_2$In compounds except *o*-Eu$_2$In (*f*=6), *o*-Yb$_2$In (*f*=13), and *o*-Y$_2$In (no *f* states) are very small ($\delta_r$<0.15). Interestingly, in case of Eu$_2$In (*f*=6) and Yb$_2$In (*f*=13), the restricted charge transfer from d-shell to f-shell due to *f*-shell closure causes charge depletion and leads to significantly reduced charge sharing in hexagonal phase. Thus, to stabilize itself, the atomic lattice of these compounds must distort locally to share required electronic charge and in doing so, both Eu$_2$In and Yb$_2$In crossed the $\delta_r$ threshold ($\delta_r$> 1.0; **Fig. 8a**). The increased chemical potential due to enhanced charge sharing leads to higher strain, i.e., large $\delta_r$, that provides enough energy to cross the energy barrier that stabilizes the orthorhombic phase, and form the necessary bonds shown in **SI Table 3-5**. This connection between 4*f*-electron configuration and energetics of lattice distortion provides fundamental insights into why the ground state crystal symmetry of RE$_2$In with *f*=6 and 13 is orthorhombic while all other RE$_2$In show hexagonal ground state. Interestingly, this shows that one can predict the possibility of PT and symmetry change in rare-earth compounds by assessing $\delta_r$ quantitatively.

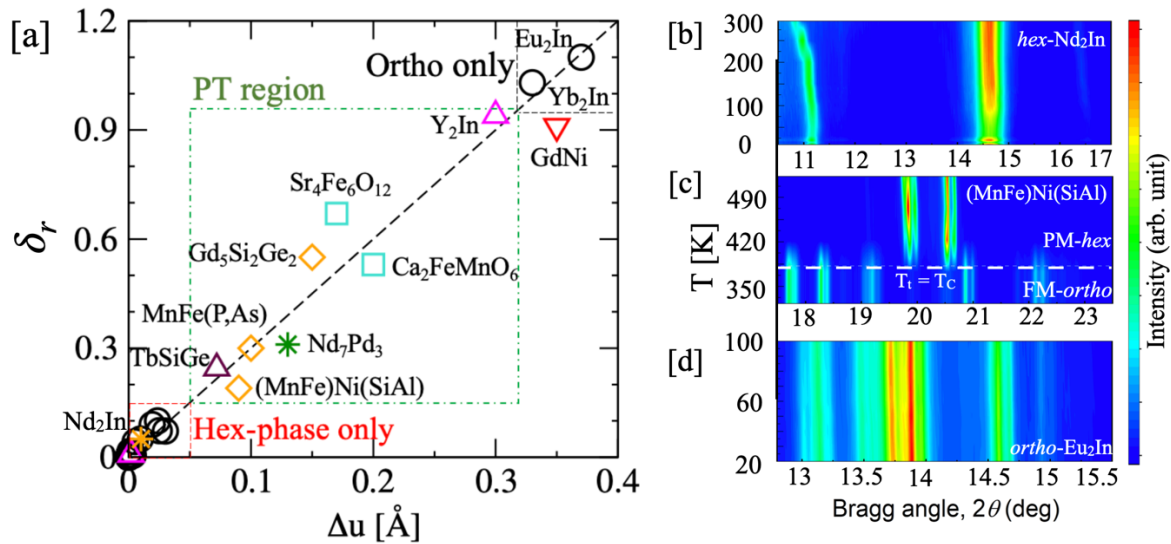

**Figure 8**. (a) DFT calculated $\delta_r$ vs vector norm of the atomic displacement, i.e., distortion, in various phases such as hexagonal (h) and orthorhombic (o) phases of various material classes. We identified three distortion regions as low ($\delta_r$<0.15), mid (0.15<$\delta_r$< 1.00), and high ($\delta_r$>1.0), each representing the stability of *hex* phase, phase transformation (e.g., *hex-ortho*), and *ortho* phase, respectively. (b-d) The experimental validation (compiled literature data) of temperature-dependent phase stability through X-ray powder diffraction on (top-to-bottom): Nd$_2$In (*hex* only; low $\delta_r$) [**38**], (Mn,Fe)Ni(Si,Al) (*hex-ortho* PT; mid $\delta_r$) [**41**], and Eu$_2$In (*ortho* only; high $\delta_r$) [**35,37**].

As discussed in case of bcc→hcp PT (**Fig. 2**), a high distortion level was seen as a key reason for martensitic PT. This hypothesis seems to connect very well within RE$_2$In magnetic compounds where less-symmetric orthorhombic ground states show very high $\delta_r$ (>1.0) while compounds that form hexagonal phase remain nearly undistorted ($\delta_r$<0.15; $\Delta u$<0.05). Notably, it is expected that the distortion will disappear or become smaller at elevated temperatures. Thus, logically the *hex*-RE$_2$In compounds with very small distortion would not change to orthorhombic phase, and therefore, temperature dependent *hex-to-ortho* PT is not plausible in such cases. Similarly, the orthorhombic Eu$_2$In and Yb$_2$In have very large $\delta_r$ and large distortion would remain intact to maintain effective charge sharing required for structural stability of orthorhombic phase. Clearly, even at higher temperatures, the high $\delta_r$ compounds such as Eu$_2$In and Yb$_2$In (see **Fig. 8a**) would not reduce to such a lower limit that may permit atomic shuffling or rearrangement into hexagonal phase (see **Ref. 68**), which precludes the occurrence of *hex-to-ortho* PT. Thus, from the foregoing discussion, one can draw two important inferences. Firstly, the $\delta_r$ is a key parameter to determine ground state crystal structure for RE$_2$In, and secondly, RE$_2$In compounds with either low $\delta_r$ (<0.15; hexagonal) or high $\delta_r$ (>1; orthorhombic) would show symmetry invariance with temperature, and for them structural transition is not favorable. While some of the RE$_2$In (Eu$_2$In, Pr$_2$In, Nd$_2$In) show first-order magnetoelastic transition, their crystal symmetry remains unchanged irrespective of temperature range. These inferences are validated via experimental studies as shown in **Fig. 8b,d**.

Following this, we turned our focus to more generalized cases, i.e., beyond RE$_2$In type compounds. For example, the non-centrosymmetric hexagonal Sm$_7$Pd$_3$ is one of rare materials showing an unusual magnetic ground-state in combination with the extraordinarily large magnetocrystalline anisotropy and nearly zero saturation magnetization [**43**]. In analogy with RE$_2$In, the much smaller $\delta_r$ (~0.03) values in *hex*-Sm$_7$Pd$_3$ suggests no possibility of structural PT. Recently concluded experiments [**43**] also show no temperature dependent structural change in *hex*-Sm$_7$Pd$_3$. On the other hand, its sister compound, i.e., Nd$_7$Pd$_3$, exhibits higher $\delta_r$ (0.31), which falls in PT region of **Fig. 8a**. Notably, recent experiments also confirm monoclinic to hexagonal PT in Nd$_7$Pd$_3$ [**71**]. Following this, we also explored few other material classes such as GdNi ($\delta_r$=1.05) and PrMnO$_3$ ($\delta_r$=0.97) that show a very high degree of local lattice distortion, and interestingly, in experiments, both the compounds show an orthorhombic ground state [**72,73**], which is consistent with our hypothesis that materials with very high or low $\delta_r$ will not exhibit any PT behavior.

Notably, in mid-to-high distortion region, the orthorhombic phase is the ground state of most of the examined compounds. Apart from RE$_2$In (first-order magnetic PT [**35**]), we calculated $\delta_r$ values for some key magneto-caloric materials such as Gd$_5$Si$_2$Ge$_2$ (giant magneto-caloric material) [**3**], (Mn,Fe)Ni(Si,Al) (both magnetocaloric and shape-memory compound) [**41**], MnFe(P,As) [**42**], and other phase changing materials that fall within PT regions shown in **Fig. 8**. Our results clearly imply that $\delta_r$ can be used as a generalized parameter to not only predict whether the ground state of a compound is hexagonal or orthorhombic but also whether any temperature induced structural PT is possible or not. As marked by three regions in **Fig. 8a**, any inorganic compound with $\delta_r$ in region (a; $\delta_r$<0.15) or (c; $\delta_r$>1.0) would show a stable ground state, i.e., no associated structural PT [**74,75**]. Conversely, we expect a structural transition for any inorganic compound with $\delta_r$ in the region (b; 0.15 <$\delta_r$<1.0). We established the generality of our approach by extending our predictions to cases beyond hexagonal (Ni$_2$In-type) to orthorhombic (Co$_2$Si-type) PT as shown in **Fig. 8**. In the past, some experimental groups have provided discussion around lattice distortion driven tetragonal to orthorhombic PT in superconducting materials [**16,17**]. These reports are suggestive of the fact that the proposed structural descriptor has a fundamental connection to underlying structure, which can be used for reliable prediction of PT in novel functional materials.

Despite fundamental advantage of our approach, we would like to point to possible material classes such as fully random or disordered materials where users might want to be little cautious in using $\delta_r$ to predict PT. In fully random or disordered systems, based on our experience, we suggest adding one extra step while connecting PT with structural PT, i.e., full thermodynamic phase stability assessment of unknown compounds or compositions in desired phases of inorganic solids including mixed rare-earth compounds [**76,77**], along with evaluation the $\delta_r$ parameter to predict possible structural PT. The underlying reason for adding extra screening step arises from possibility of several local configurations [**78**] that needs to be properly considered in PT assessment.

Furthermore, we also want to emphasize the numerical advantage of $\delta_{total}^{DFT}$ ($\sqrt{\delta_r^2 + \delta_\theta^2}$,) based scheme compared to other approaches, such as Monte Carlo method or full thermodynamic assessment using DFT. For example, the $\delta_{total}^{DFT}$ evaluation requires an "experimental reference lattice" and "fully-relaxed" (DFT) unit-cell, which is a one-step calculation followed by postprocessing of $\delta_{total}^{DFT}$ parameter as defined in method section. In this work, we generally used 6 to 24 atoms per unit-cell of intermetallic compounds each requiring 5-10 minutes of the compute time for full structural optimization (volume and atomic positions) depending on numerical complexity arising from magnetism, local correlation, and/or cell size. The post-processing time requires a fraction of second to evaluate final $\delta_{total}^{DFT}$ parameter needed to predict possible PT. On the other hand, the prediction of competing phase stabilities within DFT requires three step DFT calculation: (i) full structural relaxation, (ii) self-consistent energy calculation of competing crystallographic and magnetic phases, and (iii) post-processing of formation enthalpies and phonon stability. Notably, this requires 2.5x compute time (25-40 min without phonons; 3-5 hours including phonons) compared to one-step DFT evaluation of $\delta_{total}^{DFT}$. On the other hand, in Monte-Carlo calculations, we begin with experimental or DFT-linked thermodynamic functions or parameters as input to simulate phase-transition behavior, where parameter evaluation already requires 25-40 min as explained above, before we could start MC simulations. Depending on chemical and magnetic complexity, we need hours of equilibration and structure optimization at desired temperature, i.e., order of magnitude more time compared to $\delta_{total}^{DFT}$ evaluation.

For readers clarity and avoid any confusion between symmetry-mode analysis (X-Ray diffraction) and local lattice distortion, we want to emphasize that symmetry-mode analysis and $\delta_{total}^{DFT}$ represent two distinct approaches to studying structural changes in materials. Here, $\delta_{total}^{DFT}$ examines deviations at the atomic or nanoscale, capturing local heterogeneities such as bond length and angle variations caused by change in "non-ideal" Wycoff positions, crystal phase, local environment, and/or chemical heterogeneities etc. Symmetry-mode analysis focuses on group-theoretical and global symmetry considerations, using the periodicity and symmetry of the unit cell to decompose structural distortions into symmetry-adapted modes. This work shows a successful attempt connect symmetry changes during structural phase transition with $\delta_{total}^{DFT}$ including the effect of collective atomic movements, provides crucial insights into localized quantum mechanical effects.

Finally, we want to emphasize that despite being a DFT-based approach, we show that the $\delta_{total}^{DFT}$ parameter is numerically inexpensive to estimate, therefore, much faster to execute, which gives an edge over other methods in assessing new compounds for possibility of PTs. This quantitative method for assessing lattice distortions lays the groundwork for future studies on phase transitions, stability, and functional properties in complex materials.

**Summary**

In this communication, we presented a new structural descriptor, distortion parameter, to predict PT in inorganic solids. We have shown that DFT-derived local atomic strain correlates very well with observed phase transformations, which has a strong effect on structural (phonons) stability and electronic structure of inorganic solids. The origin of structural phase transformation lies in charge induced atomic displacements that are suppressed in the high-symmetry structure, e.g. hexagonal, phase while low symmetry, e.g. orthorhombic, phase of same material shows preference for higher distortion. The descriptor was linked to charge-driven atomic distortions as well as to structural and magnetic phase transitions observed in a series of RE$_2$In compounds. We used the $\delta_{total}^{DFT}$ to predict compositional structural changes in two new pseudo-binary systems, i.e., (Yb$_{1-x}$Er$_x$)$_2$In and Gd$_2$(In$_{1-x}$Al$_x$), and validated them by our X-ray powder diffraction experiments. Finally, by underpinning the fundamental physics of phase transformations in RE$_2$In compounds, we were able to generalize this knowledge to other compounds with potential structural transformations that involve a change in local crystal symmetry.

Our ability to link phase transitions with $\delta_{total}^{DFT}$ and underlying charge mechanism shows that the degree of atomic distortion can be systematically tuned via change in elemental chemistry, defects, or other external parameters to control phase change behavior of complex inorganic solids. Finally, we also rationalize the numerical advantage of DFT driven lattice and angular distortion-based approach by providing a qualitative compute time comparison with other methods such as Monte-Carlo or DFT approaches that involve more numerically intensive thermodynamic processes and, therefore, are numerically inefficient for high-throughput assessment of structural phase transition in inorganic solids.

**Acknowledgement**

We would like to thank Prof. Paul C. Canfield and Prof. Duane D Johnson for fruitful discussion. The work at Ames National Laboratory was supported by the U.S. DOE, Office of Science, Basic Energy Sciences (BES), Materials Science & Engineering Division. Ames National Laboratory is operated by Iowa State University under contract DE-AC02-07CH11358.

**Author Contribution**

PS conceptualized the idea, performed DFT calculations, analyzed the theoretical results, and wrote the first draft of the manuscript. AB and YM proposed the problem. AT and YM performed experiments. PS, AB, and YM contributed to analysis, discussion, and writing of the final manuscript.

**References**

1. S.B. Roy, First-order magneto-structural phase transition and associated multi-functional properties in magnetic solids, J. Phys. Condens. Mat. **25,** 183201 (2013).
2. Z. Sun *et al.* Ultrahigh pyroelectric figures of merit associated with distinct bistable dielectric phase transition in a new molecular compound: di-n-butylaminium trifluoroacetate, Adv. Mater. **27**, 4795-4801 (2015).
3. V. K. Pecharsky, and K. A. Gschneidner, Giant Magnetocaloric Effect in Gd$_5$(Si$_2$Ge$_2$), Phys. Rev. Lett. **78**, 4494 (1997).
4. Y. Tokura, Colossal Magnetoresistive Oxides, Gordon and Breach Science Publishers, The Netherlands,

2000.
5. M. R. Ibarra and P. A. Algarabel, Giant volume magnetostriction in the FeRh alloy, Phys. Rev. B **50**, 4196 (1994).
6. N. A. Spaldin, and R. Ramesh, Advances in magnetoelectric multiferroics, Nature Materials **18**, 2023-212 (2019).
7. B. Neese et al. Large electrocaloric effect in ferroelectric polymers near room temperature, Science **321**, 821-823 (2008).
8. J. Cui, Y. Chu, O. Famodu *et al*. Combinatorial search of thermoelastic shape-memory alloys with extremely small hysteresis width, Nature Mater **5**, 286-290 (2006).
9. D. X. Liu, H.L. Zhu, W.X. Zhang, and X.M. Chen, Nonlinear optical glass-ceramic from a new polar phase-transition organic-inorganic hybrid crystal, Angew. Chem. Int. Ed. **62**, e202218902 (2023).
10. C. Qiu et al. Transparent ferroelectric crystals with ultrahigh piezoelectricity, Nature **577**, 350-354 (2020).
11. Y. Hu et al. Ferroelastic-switching-driven large shear strain and piezoelectricity in a hybrid ferroelectric, Nat. Mater. **20**, 612-617 (2021).
12. K. Ding et al. Superior ferroelectricity and nonlinear optical response in a hybrid germanium iodide hexagonal perovskite, Nat. Commun. **14**, 2863 (2023).
13. L. Tang et al. Giant near-room-temperature pyroelectric figures-of-merit originating from unusual dielectric bistability of two-dimensional perovskite ferroelectric crystals, Chem. Mater. **34**, 8898-8904 (2022).
14. C.-D. Liu et al. Spectrally selective polarization-sensitive photodetection based on a 1D lead-free hybrid perovskite ferroelectric, ACS Mater. Lett. **5**, 1974–1981 (2023).
15. V.L. Deringer, R. Dronskowski, and M. Wuttig, Microscopic Complexity in Phase-Change Materials and its Role for Applications, Adv. Funct. Mater. **25**, 6343-6359 (2015).
16. Q. Si, R. Yu, and E. Abrahams, High-temperature super- conductivity in iron pnictides and chalcogenides, Nat. Rev. Mater. **1**, 16017 (2016).
17. A. E. Böhmer *et al.* Effect of Biaxial Strain on the Phase Transitions of Ca(Fe$_{1-x}$Co$_x$)$_2$As$_2$, Phys. Rev. Lett. **118**, 107002 (2017).
18. A. Kitanovski, Energy Applications of Magnetocaloric Materials Adv. Energy Mater. **10**, 190374 (2020).
19. I. Hughes, M. Dane, A. Ernst, W. Hergert, M. Luders, J. Poulter, J. B. Staunton, A. Svane, Z. Szotek, and W. M. Temmerman, Lanthanide contraction and magnetism in the heavy rare earth elements, Nature **446**, 650-653 (2007).
20. K. Binder, E. Luijten, M. Müller, N. B. Wilding, and H.W.J. Blöte, Monte Carlo investigations of phase transitions: status and perspectives, Physica A **281**, 112-128 (2000).
21. J. Ding, H.-K. Tang, and W. C. Yu, Rapid detection of phase transitions from Monte Carlo samples before equilibrium, SciPost Phys. **13**, 057 (2022).
22. C. Niu, Y. Rao, W. Windl, and M. Ghazisaeidi, Multi-cell Monte Carlo method for phase prediction, npj Comput Mater. **5**, 120 (2019).
23. C.J. Bartel, Review of computational approaches to predict the thermodynamic stability of inorganic solids. J Mater Sci **57**, 10475–10498 (2022).
24. Y.F. Ye, C.T. Liu, and Y. Yang, A geometric model for intrinsic residual strain and phase stability in high entropy alloys, Acta Mater. **94,** 152-161(2015).

25. X. Zhou, A. Tehranchi, and W. A. Curtin, Mechanism and Prediction of Hydrogen Embrittlement in fcc Stainless Steels and High Entropy Alloys, Phys. Rev. Lett. **127**, 175501 (2021).
26. S.F. Pugh, Relations between elastic moduli and the plastic properties of polycrystalline pure metals. Lond. Edinb. Dublin Philos. Mag. **45**, 8230843 (1954).
27. P. Singh, B. Vella, G. Ouyang, N. Argibay, J. Cui, R. Arroyave, and D.D. Johnson, A ductility metric for refractory-based multi-principal element alloy, Acta Mater. **257**, 1-15 (2023).
28. D.D. Johnson, P. Singh, A.V. Smirnov, and N. Argibay, Universal maximum strength of solid metals and alloy, Phys. Rev. Lett. **130**, 166101 (2023).
29. M. Chandross, and N. Argibay, Ultimate strength of metals, Phys. Rev. Lett. **124** (12), 125501 (2020).
30. H. Song *et al.* Local lattice distortion in high-entropy alloys, Phys. Rev. Mater. **1**, 023404 (2017).
31. G. Kresse, and J. Hafner, Ab initio molecular dynamics for liquid metals, Phys. Rev. B **47**, 558-561 (1993).
32. G. Kresse, and D. Joubert, From ultrasoft pseudopotentials to the projector augmented-wave method, Phys. Rev. B **59**, 1758-1775 (1999).
33. A. Najev *et al.* Uniaxial Strain Control of Bulk Ferromagnetism in Rare-Earth Titanates, Phys. Rev. Lett. **128**, 167201 (2022).
34. E. Svanidze, C. Georgen, A. M. Hallas, Q. Huang, J. M. Santiago, J. W. Lynn, and E. Morosan, Band Jahn-Teller structural phase transition in $Y_2In$, Phys. Rev. B **97**, 054111 (2018).
35. F. Guillou, A.K. Pathak, D. Paudyal, Y. Mudryk, F. Wilhelm, A. Rogalev, and V. K. Pecharsky, Non-hysteretic first-order phase transition with large latent heat and giant low-field magnetocaloric effect, Nat Commun. **9**, 2925 (2018).
36. E. Mendive-Tapia, D. Paudyal, L. Petit, and J. B. Staunton, First-order ferromagnetic transitions of lanthanide local moments in divalent compounds: An itinerant electron positive feedback mechanism and Fermi surface topological change, Phys. Rev. B **101**, 174437 (2020).
37. F. Guillou, H. Yibole, R. Hamane, V. Hardy, Y. B. Sun, J. J. Zhao, Y. Mudryk, and V. K. Pecharsky, Crystal structure and physical properties of $Yb_2In$ and $Eu_{2-x}Yb_xIn$ alloys, Phys. Rev. Mater. **4**, 104402 (2020).
38. A. Biswas, R. K. Chouhan, A. Thayer, Y. Mudryk, I. Z. Hlova, O. Dolotko, and V. K. Pecharsky, Unusual first-order magnetic phase transition and large magnetocaloric effect in $Nd_2In$, Phys. Rev. Mater. **6**, 114406 (2022).
39. W. Cui, G. Yao, S. sun, Q. Wang, J. Zhu, and S. Yang, Unconventional metamagnetic phase transition in $R_2In$ (R=Nd, Pr) with lamda-like specific heat and nonhysteresis, J. Mater. Sci. Technol. **101**, 80 (2022).
40. A. Biswas, N.A. Zarkevich, A.K. Pathak, O. Dolotko, I.Z. Hlova, A.V. Smirnov, Y. Mudryk, D.D. Johnson, and V.K. Pecharsky, First-order magnetic phase transition in $Pr_2In$ with negligible thermomagnetic hysteresis, Phys. Rev. B **101**, 224402 (2020).
41. A. Biswas, A. K. Pathak, N.A. Zarkevich, L. Xubo, Y. Mudryk, V. Balema, D.D. Johnson, and V. K. Pecharsky, Designed materials with the giant magnetocaloric effect near room temperature, Acta Mater **180**, 341-348 (2018).
42. O. Tegus, E. Brück, KH.J. Buschow, and F. R. de Boer, Transition-metal-based magnetic refrigerants for room-temperature applications. Nature **415**, 150–152 (2002).
43. A. Biswas, R. K. Chouhan, O. Dolotko, P. Manfrinetti, S. Lapidus, D. L. Schlagel, and Y. Mudryk. Exceptional magnetic and magnetoelastic behavior of rare-earth non-centrosymmetric $Sm_7Pd_3$, Acta

Mater. **265**, 119630 (2024).
44. A. Palenzona, The crystal structure and lattice constants of RE$_2$In and some RE$_5$In$_3$ compounds, J. Less-Common Met. **16**, 379-384 (1968).
45. H.J., Monkhorst, and J.D. Pack, Special points for Brillouin-zone integrations, Phys. Rev. B **13**, 5188-5192 (1976).
46. J.P. Perdew, K. Burke, and M. Ernzerhof, Generalized gradient approximation made simple, Phys. Rev. Lett. **77**, 3865-3868 (1996).
47. P. Singh, M.K. Harbola, M. Hemanadhan, A. Mookerjee, and D.D. Johnson, Better band gaps with asymptotically corrected local exchange potentials, Phys Rev B 93, 085204 (2016).
48. P. Singh, M.K. Harbola, B. Sanyal, and A. Mookerjee, Accurate determination of band gaps within density functional formalism, Phys Rev B 87, 235110 (2013).
49. S.L. Dudarev, G.A. Botton, S.Y. Savrasov, C.J. Humphreys, and A.P. Sutton, Electron-energy-loss spectra and the structural stability of nickel oxide: An LSDA+U study, Phys. Rev. B **57**, 1505 (1998).
50. M. Topsakal, and R.M. Wentzcovitch, Accurate projected augmented wave (PAW) datasets for rare-earth elements (RE = La–Lu), Comput Mater Sci **95**, 263-270 (2014).
51. Y. Zhang, Y. J. Zhou, J.P. Lin, G.L. Chen, and P.K. Liaw, Solid-solution phase formation rules for multi-component alloys, Adv. Eng. Mater. **10**, 534-538 (2008).
52. N. M. Fortunato et al. High-Throughput Screening of All-d-Metal Heusler Alloys for Magnetocaloric Applications, Chem. Mater. **36**, 6765–6776 (2024).
53. S. Baroni, S. de Gironcoli, A. Dal Corso, and P. Gianozzi, Phonons and related crystal properties from density-functional perturbation theory, Rev. Mod. Phys. **73**, 515-562 (2001).
54. L. Chaput, A. Togo, I. Tanaka, and G. Hug, Phonon-phonon interactions in transition metals, Phys. Rev. B **84**, 094302 (2011).
55. B. Lee, and X. Gonze, Ab initio calculation of the thermodynamic properties and atomic temperature factors of SiO$_2$ α-quartz and stishovite, Phys. Rev. B **51**, 8610 (1995).
56. K. Parlinski, Z.Q. Li, and Y. Kawazoe, First-Principles Determination of the Soft Mode in Cubic ZrO$_2$, Phys. Rev. Lett. **78**, 4063 (1997).
57. A. Togo, and I. Tanaka, First principles phonon calculations in materials science, Scripta Mater. **108**, 1 (2015).
58. W.G. Burgers, On the process of transition of the cubic-body-centered modification into the hexagonal-close-packed modification of zirconium, Physica **1**, 561-586 (1934).
59. J.S. Bowles, and J.K. MacKenzie, The crystallography of martensite transformations—IV body-centred cubic to orthorhombic transformations, Acta Metall. **5**, 137-149 (1957).
60. P. Gaunt, and J.W. Christian, The crystallography of the β-α transformation in zirconium and in two titanium-molybdenum alloys, Acta Metall. **7**, 534-543 (1959).
61. S. Banerjee, G.K. Dey, D. Srivastava, and S. Ranganathan, Plate-Shaped Transformation Products in Zirconium-Base Alloys, Metall. Mater. Trans **28A**, 2201-2216 (1997).
62. P. Singh, A.V. Smirnov, and D.D. Johnson, Ta-Nb-Mo-W refractory high-entropy alloys: anomalous ordering behavior and its intriguing electronic origin, Phys. Rev. Mater. **2** (5), 055004 (2018).
63. P. Singh, T. Del Rose, and Y. Mudryk, Effect of disorder on thermodynamic instability of binary rare-earth – nickel – palladium compounds, Acta Mater. **238**, 118205 (2022).
64. Y.J. Shi, Y.L. Du, G. Chen, and G.L. Chen, A First principle study on phase stability and electronic

structure of YCu, Physics Letters A **368**, 495-498 (2007).
65. S. Asano, and S. Ishida, Band Jahn-Teller Effect in LaCd, J. Phys. Soc. Jpn. **54**, 4241-4245 (1985).
66. K.A. Gschneidner, Y. Mudryk, and V.K. Pecharsky, On the nature of the magnetocaloric effect of the first-order magnetostructural transition, Scripta Mater. **67**, 572-577 (2012).
67. A. Fujita, S. Fujieda, Y. Hasegawa, and K. Fukamichi, Itinerant-electron metamagnetic transition and large magnetocaloric effects in La(Fe$_x$Si$_{1-x}$)$_{13}$ compounds and their hydrides, Phys. Rev. B **67**, 104416 (2003).
68. F. Guillou, G. Porcari, H. Yibole, N. van Dijk, and E. Brück, Taming the first-order transition in giant magnetocaloric materials, Adv. Mater. **26**, 2671-2675 (2014).
69. Y. Yu et al. Enhanced magnetocaloric effects in Gd$_2$In$_{1-x}$Al$_x$ (0.4 ≤ x ≤ 1) system by the hysteresis-free metamagnetism, Journal of Magnetism and Magnetic Materials **524**, 167648 (2021).
70. C.M. Wayman, Introduction to the crystallography of martensitic transformations, Macmillan, New York, 1964.
71. Y. Mudryk *et al.* Magnetostructural behavior in the non-centrosymmetric compound Nd$_7$Pd$_3$, J. Phys.: Condens. Matter **31**, 265801 (2019).
72. A. P. Alho *et al.* Fathoming the anisotropic magnetoelasticity and magnetocaloric effect in GdNi, Phys. Rev. B **106**, 184403 (2022).
73. J. Hemberger, M. Brando, R. Wehn, V. Yu. Ivanov, A. A. Mukhin, A. M. Balbashov, and A. Loidl, Magnetic properties and specific heat of RMnO$_3$ (R = Pr, Nd), Phys. Rev. B **69**, 064418 (2004).
74. C.-D.-L. Cruz *et al.* Lattice Distortion and Magnetic Quantum Phase Transition in CeFeAs$_{1-x}$P$_x$O, Phys. Rev. Lett. **104**, 017204 (2010).
75. A. Bianconi, N. L. Saini, A. Lanzara, M. Missori, and T. Rossetti, Determination of the Local Lattice Distortions in the CuO$_2$ Plane of La$_{1.85}$Sr$_{0.15}$CuO$_4$, Phys. Rev. Lett. **76**, (1996).
76. P. Singh, Role of spin-orbit coupling on crystal-field splitting and phase-stability of rare-earth based layered intermetallic, Scripta Materialia **236**, 115644 (2023).
77. P. Singh, T. Del Rose, G. Vazquez, R. Arroyave, and Y. Mudryk, Machine-learning enabled thermodynamic model for the design of new rare-earth compounds, Acta Materialia **229**, 117759 (2022).
78. R. Singh, A. Sharma, P. Singh, G. Balasubramanian, and D.D. Johnson, Accelerating computational modeling and design of high-entropy alloys, Nature Computational Science **1** (1), 54-61 (2021).